# A Field-Weighted model for Surface Layer Characterization using Single Channel Intensity Interrogation SPR

Zhiying Chen, Zihao Luo, Changsen Sun, Dmitry Kiesewetter, Sergey Krivosheev, Sergey Magazinov, Victor Malyugin, Xue Han *

**ABSTRACT:** To address the difficulty of characterizing the surface layer rigorously, especially the thickness and refractive index (RI) in surface plasmon resonance (SPR) technology, we propose a field-weighted analysis method. This approach enables simultaneous quantitative determination of RI for the bulk solution and the surface layer. This study utilizes the aluminum-based Kretschmann structure with the intensity interrogation technique. We construct the field-weighted model governed by the evanescent field penetration depth to decompose the SPR reflected intensity into the bulk and surface responses. Experiments are conducted using bovine serum albumin (BSA) solution to form a surface adsorbed protein layer, and different concentrations of BSA are tested. Results show that the separated surface response fits well with the Langmuir formula, representing a significant improvement over the untreated SPR signal. The bulk and surface responses are then incorporated into the field-weighted model to determine the RI values of the bulk BSA solution and the surface adsorbed BSA layer at various concentrations. The experimental results of BSA solution match the Abbe refractometer measurements with a maximum error 0.0004 in RI, while the results of the adsorbed BSA layer, both the RI and thickness, aligned well with reported parameters for a single BSA layer. This method eliminates the stage rotation in the common angular interrogation SPR technique and complicated optical design and nano-fabrication in the nano-optics sensing schemes, making it suitable for compact, low-cost SPR platforms for practical applications needing surface layer characterization.

Surface plasmon resonance (SPR) is a physical phenomenon that occurs when an incident light beam excites the coherent oscillation of free electrons at the interface between a metal and a dielectric medium at the resonance condition, i.e. frequency and momentum matching[1][3]. Around this resonance condition, an evanescent field is generated and decays into the dielectric medium. Due to the exponential nature of this evanescent field, the surface changes, e.g. molecular binding, conformational changes, and modifications to the surface layer, can be detected by SPR techniques with extremely high sensitivities[4],[5]. Therefore, it has become one of the widely applied optical techniques in the surface process monitoring, playing an indispensable role in studies such as antigen-antibody recognition[6],[7], drug target affinity assays[-8-9], and surface functionalization analysis[10][11].

However, the evanescent fields of SPR sensors can penetrate dielectric media to depths of hundreds of nanometers[12], far exceeding the characteristic thickness of most target analytes—such as proteins measuring from 2 nm to 10 nm in size[13][14]. Therefore, even minor variations in the bulk solution which covers the majority of the evanescent field can dominate the SPR response and bury the relatively small contributions from nanoscale surface layers[15],[16].

This issue is particularly acute for ultrathin films such as adsorbed proteins, grafted polymers, and lipid membranes, whose optical properties are confined to a range of just a few nanometers. However, for traditional SPR sensors, whether based on angle[17],[18], wavelength[19][20] or intensity[21] interrogations, only a single scalar measured value, such as resonance angle/wavelength displacement or intensity change, is obtained. This value inherently consists of the changes from the surface layer and the bulk solution.

To factor out the bulk solution contribution in the SPR sensing for the precise surface sensing, researchers have proposed various correction strategies. The Chinowsky group proposed that the total internal reflection (TIR) angle was primarily determined by the bulk solution, therefore this critical angle can be used to determine the bulk solution RI $n_b$ with Snell's equation. The surface response was characterized by $n_{surface} = n_{SPR} - n_b$ where $n_{SPR}$ was the RI determined by the SPR measurement and $n_{surface}$ represented the RI of the surface layer[22][23]. Without separating the volume weights for the surface layer and the bulk solution, this method is underestimated the RI value of the surface layer. Furthermore, Svirelis et al. advanced a correction method from the simple subtraction method to a field-weighted model based on the field attenuation exponential since the bulk solution actually covered a partial of the evanescent field. The bulk response was calculated as the production of its corresponding weight and the original SPR signal[24]. While Svirelis's method rigorously eliminates the bulk influence, it does not offer a direct mechanism to quantify RI values of the bulk and surface components independently. In addition to utilizing TIR critical angle to extract $n_b$, some studies employ other modes within optical nano-structures to separate the surface and the bulk responses. Zeng et al. designed interferometric ring-hole arrays and each unit contained a few concentric ring apertures with varied radii to generate different surface plasmon polariton (SPP) waves propagating toward the central hole. These SPP waves interfered and peaks and valleys were resulted in the transmission spectra. Each valley and peak was featured specific surface and bulk sensitivities. When changes happened to the surface or/and the bulk medium, the interference peak and valley location shifted. The changes of the surface layer thickness ($\Delta d_s$) and the bulk RI ($\Delta n_b$) were calculated by using two interference peaks/valleys through formula manipulation[25]. Liu et al. exploited the bulk-specific sensitive near-cutoff (NC) mode in the tilted fiber Bragg grating-SPR (TFBG-SPR) to factor out the bulk solution fluctuations. Through a pre-calibrated relationship between the NC and SPR resonances, the bulk RI contribution was calculated and deducted to extract the surface signal[26]. The approaches proposed by Liu and Zeng require complex nanofabrication and entail significant costs, limiting their practical applicability. The RI of a surface layer carries core information such as the film density, hydration state, packing structure, and intermolecular interactions[27]. These factors collectively determine the functional properties of biomolecular interfaces[28] and hold significant implications for biomedical applications including drug development and vaccine production[29]. Consequently, establishing a concise model capable of accurately decoupling and quantifying the bulk and surface RI variations remains a critical challenge.

In this work, we utilize an aluminum-based Kretschmann configuration and establish a field-weighted model governed by the evanescent field penetration depth to decompose the total optical response to extract RI values of the surface and bulk components with the SPR intensity interrogation technique. Multiple injection (protein solution) - flushing (baseline solution) steps are used to form a dense bovine serum albumin (BSA) single layer. We separate the surface and bulk responses by comparing the SPR responses before and after saline baseline flush. Based on the isolated surface response, a dense monolayer of BSA is confirmed by the Langmuir analysis which accurately reflects the BSA adsorption behavior. Finally, the separated surface and bulk responses are substituted into the filed-weighted model to calculate the RI values of the bulk BSA solution and the BSA adsorbed layer at different BSA concentrations. These results match measurements from an Abbe refractometer and reported BSA layer in literature. The validity of the assigned thickness of the adsorption layer is also evaluated which matches literature report. This field-weighted model enables quantitative characterization of protein adsorbed layers with RI and thickness. This proposed SPR intensity interrogation protocol eliminates the need of the angle scanning, multi-channel detection, or complex optical structures, and facilitates a compact and low-cost SPR sensing platform for surface layer characterization.

■ **MATERIALS AND METHODS**

SPR excitation based on aluminum (Al) films shows high tunability and application potential from the deep UV to the near infrared regimes[30]. And our previous work has demonstrated a record high sensitivity based on aluminum-based angular interrogation SPR using the blue light[31]. Al films are prepared using the thermal evaporation technique (TECHNOL ZHD300 high-vacuum resistive evaporation coating system) and the detailed procedure is described in the supporting information (S1).

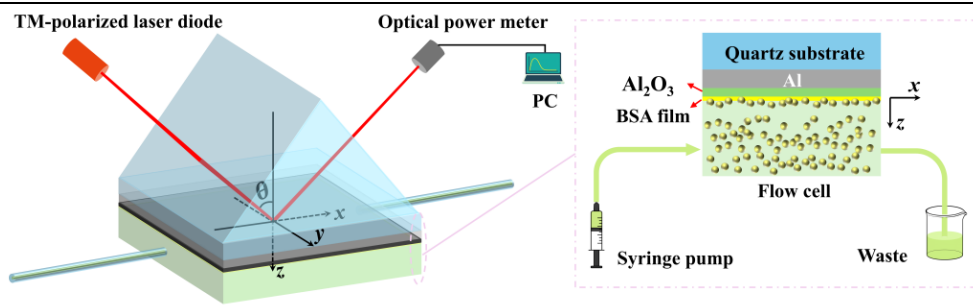

**Fig. 1.** Schematic illustration of the Kretschmann coupling and the multilayer SPR configuration. The surface BSA film and bulk BSA region are illustrated.

Studies have demonstrated that BSA exhibits excellent adsorption properties on oxide surfaces[32]. In our experiment, the saline (0.9% NaCl) is used as the solvent for BSA, and 20, 40, 60, 100, and 140 μM BSA are prepared. All solutions are freshly prepared prior to the experiment. The refractive indices of the DI water, saline, and BSA solutions at various concentrations are measured using an Abbe refractometer. The RI values of DI water and saline solution are used to determine the sensitivity of the SPR intensity interrogation.

The SPR experiment is conducted using the Kretschmann configuration with an equilateral prism featuring a RI of 1.4588 at 635 nm. The excitation source is a semiconductor laser emitting at 635 nm wavelength. Reflected light intensity is measured using a power meter from Thorlabs. BSA molecule does not absorb 635 nm light.

An angular interrogation is performed in the DI water solution, and the details of the procedure is described in the supporting information (S1). The experimental and theoretical calculation of the angular spectra are compared to demonstrate the validity of the parameters, i.e. thickness and real and imaginary values of the RI, of the Al thin film. The theoretical calculation method is detailed in the supporting information (Fig. S2). Based on the angular spectrum, a suitable angle is selected as the fixed working angle for the intensity interrogation. Subsequently, the power meter is fixed at the corresponding reflection angle. Solutions of varying concentrations of BSA are sequentially introduced into the flow cell and saline solution is used to flush the floating proteins after each BSA solution injection. The reflected light intensity is recorded in real-time.

This work selected Al as the SPR sensing substrate. On one hand, Al films naturally form a stable $Al_2O_3$ layer in air, which provides favorable surface conditions for BSA adsorption. On the other hand, simulation results indicate that aluminum-based SPR structures struggle to adopt the TIR reference channel strategy commonly used in gold-based systems. For Al films, the TIR edge closely overlaps with the SPR edge, making it difficult to effectively distinguish between the two signal types (Fig. S3). Another issue from the angular interrogation is the mechanical unstable fact from rotating the flow cell holder stage. The intensity interrogation strategy is selected for this research. Nevertheless, the broadband plasmonic response of Al films provides a unique advantage, enabling multi-wavelength SPR sensing. This property has been validated through comparative angle scans at three distinct wavelengths (Fig. S4). In this work, an Al film thickness of 13.4 nm with an $Al_2O_3$ layer thickness of approximately 4.0 nm is used. Considering the linear sensing range, the incident angle 69.3° is selected as the fixed operating angle for intensity interrogation (corresponding to the 3/4 dip position, Fig. S3).

■ **THEORY of Field-Weighted Model**

The structure of the Kretschmann configuration is illustrated in Fig. 1(a). The interface between the natural $Al_2O_3$ layer and the solution medium is defined as $z = 0$, with the positive $z$-axis pointing into the solution as in Fig. 1(b). When BSA solution is introduced, BSA adsorb onto the $Al_2O_3$ surface, and a dense single layer of BSA is formed eventually.

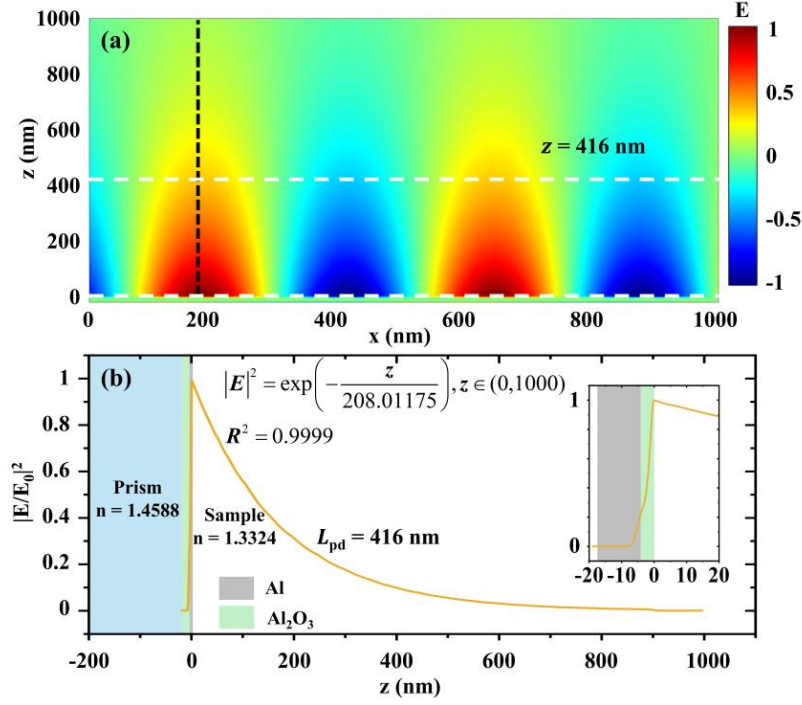

**Fig. 2.** (a) Calculated spatial distribution of the electric field intensity $|E|^2$ near the metal–dielectric interface at the operation angle. (b) Power of the normalized electric field intensity decay along the surface-normal direction, illustrating the exponential attenuation of the evanescent field and the value of the penetration depth $L_{pd}$.

The refractive index profile is stratified into two distinct regions, the surface adsorption layer ($n_s$) and the bulk solution ($n_b$) extending through the rest of the evanescent field.

The SPP field $E(z)$ exhibits exponential decay along $z$ direction into the solution medium, as Eq. (1)[12],[33],

$$\boldsymbol{E}(z) = \boldsymbol{E}(0)exp(-\frac{z}{L_{pd}}), \qquad (1)$$

where $z$ is the spatial coordinate perpendicular to the interface, $E(0)$ is the localized electric field amplitude at the interface ($z = 0$) and its magnitude is determined by SPR excitation conditions, and $L_{pd}$ is the penetration depth of the SPP. The SPR reflected intensity ($I$), measured at the operation angle of the intensity interrogation scheme, is fundamentally related to the localized electromagnetic energy density, proportional to the square of the electric field amplitude as expressed in Eq. (2).

$$|\boldsymbol{E}(z)|^2 = |\boldsymbol{E}(0)|^2 exp(-\frac{2z}{L_{pd}}), \qquad (2)$$

The spatial distribution of $E(z)$ is shown **Fig. 2(a)**. The propagating feature at the metal/solution interface of the SPP wave is observed. The evanescent field decay property is demonstrated by plotting the power of normalized $E(z)$, in **Fig. 2(b)**, along the select position on the $x$-axis as the black dashed line shown in **Fig. 2(a)**. It can be observed that the electric field is significantly amplified at the metal-dielectric interface and rapidly decays along the $z$ direction, exhibiting the characteristic features of an evanescent field. The plots in **Fig. 2(a)** and **(b)** are based on the Al film with the thickness of 13.4 nm and 4.0 nm oxide layer using a 635 nm beam at the fixed incident angle. The surface plasmon penetration depth $L_{pd}$ obtained from the fitting result is 416 nm and adopted in the experimental data analysis.

As shown in **Fig. 1(b)**, the surface layer occupies the volume from $z = 0$ to $z = d_s$, and the solution covers the rest of the evanescent field. The RI change profile can be divided into two components, the surface layer and the bulk medium:

$$\Delta n(z) = \begin{cases} \Delta n_s, & 0 < z < d_s, \\ \Delta n_b, & z \geq d_s, \end{cases} \qquad (3)$$

where $d_s$ is the effective thickness of the surface layer. BSA has the specific size around 3 nm[34],[35]. For a dense single layer of BSA molecule, the thickness of the surface absorbed layer can be assumed as 3 nm.

The SPR intensity interrogation operates within the linear region, i.e. the reflected intensity $I$ exhibiting a linear response to a RI variation. The overall change in the reflected intensity $\Delta I$ can be expressed as a spatially weighted integral of the RI perturbation as Eq. (4),

$$\Delta I = |E(0)|^2 \int_0^\infty \Delta n(z) exp(-\frac{2z}{L_{pd}}) dz, \quad (4)$$

Substituting $\Delta n$ from Eq. (3) into Eq. (4), the intensity change from the surface layer is denoted as $\Delta I_s$ and $\Delta I_b$ for the bulk solution.

$$\Delta I = \Delta I_s + \Delta I_b \\ = |E(0)|^2 [\Delta n_s \int_0^{d_s} exp(-\frac{2z}{L_{pd}}) dz + \Delta n_b \int_{d_s}^\infty exp(-\frac{2z}{L_{pd}}) dz], \quad (5)$$

Integrate over the surface layer component,

$$\Delta I_s = |E(0)|^2 \frac{L_{pd}}{2} \Delta n_s \left(1 - e^{-2d_s/L_{pd}}\right) \\ = |E(0)|^2 \frac{L_{pd}}{2} \Delta n_s w_s, \quad (6)$$

where $w_s = 1 - \exp(-\frac{2d_s}{L_{pd}})$ represents the effective weight of the SPR evanescent field within the surface layer. If $d_s = 3$ nm and $L_{pd} = 416$ nm, we obtain $w_s = 0.0143$.

Integrate over the bulk solution component,

$$\Delta I_b = |E(0)|^2 \frac{L_{pd}}{2} \Delta n_b (1 - w_s) \quad (7)$$

In this field-weighted model, the thickness of the surface layer is significantly smaller than the penetration depth of the evanescent field. At the experimental condition, the surface weighting factor $w_s$ is approximately 0.0143, while the bulk weighting factor $w_b = 1 - w_s$ is close to 1. Thus, the integral expression for the bulk response can be modified as Eq. (8):

$$\Delta I_b' = |E(0)|^2 \frac{L_{pd}}{2} \Delta n_b' \quad (8)$$

here, $\Delta I_b'$ and $\Delta n_b'$ represent the reflected intensity and RI change of the bulk solution, respectively. Eq. (8) also can be used for the scenarios only bulk change involved. For the SPR intensity interrogation, the sensing sensitivity, commonly used to transduce the RI of the bulk solution, is determined by DI water ($n_{water} = 1.3324$) and saline ($n_{saline} = 1.3343$) experimentally.

Figure 3 illustrates the systematic measurement protocol employed in this study to decouple the surface and the bulk responses. Initially, the flow cell is filled with the saline solution, with the corresponding reflected light intensity recorded as $I_0$. Subsequently, the BSA solution is injected into the flow cell (Step A). Upon reaching the adsorption equilibrium (Step B), BSA molecule form a nanoscale adsorption layer on the alumina surface, with the corresponding steady-state reflected light intensity recorded as $I_1$. Next, the saline is slowly introduced to replace the BSA solution in the sample flow cell (Step C), and the reflected light intensity at this point is recorded as $I_2$. Varying the concentration of the BSA solution, three steps A→B→C are repeated, and the corresponding changes in light intensity are recorded. Based on these measurements, the net SPR response attributed to the surface BSA layer $\Delta I_s$ equals $I_2 - I_{saline}$, while the bulk BSA solution $\Delta I_b$ equals $I_1 - I_2$. The changes in SPR reflected intensity of the surface BSA layer and bulk BSA solution are extracted.

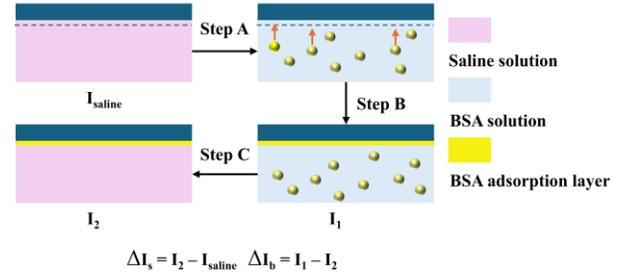

**Fig. 3.** Schematic diagram of the SPR measurement protocol employed to decouple bulk and surface responses. $I_0$, $I_1$ and $I_2$ denote the stabilized reflected intensities measured for the saline baseline, the BSA solution at equilibrium, and the post-rinse saline phase, respectively.

Based on $\Delta I_b$ signal and Eqs. (7) and (8), the RI variation in the bulk is expressed as follows,

$$\frac{\Delta I_b}{\Delta I_b'} = \frac{\Delta n_b (1 - w_s)}{\Delta n_b'} \quad (9)$$

i.e.

$$\Delta n_b = \frac{\Delta I_b}{\Delta I_b'} \frac{\Delta n_b'}{1 - w_s} \quad (10)$$

Here, $\Delta n_b'$ is the RI difference between DI water and saline which are measured by an Abbe refractometer. The precise RI of the bulk solution can thus be obtained:

$$n_b = \Delta n_b + n_{saline} \quad (11)$$

Based on $\Delta I_s$ signal, Eqs. (6) and (8), the surface RI variation can be derived as follows,

$$\frac{\Delta I_s}{\Delta I_b'} = \frac{\Delta n_s w_s}{\Delta n_b'} \quad (12)$$

i.e.

$$\Delta n_s = \frac{\Delta I_s}{\Delta I_b'} \frac{\Delta n_b'}{w_s} \quad (13)$$

The precise RI of the surface adsorption layer can thus be obtained:

$$n_s = \Delta n_s + n_{saline} \quad (14)$$

We should notice that the linear region should be roughly determined with the numerical or analytical simulation before carrying out the actual experiments.

BSA adsorption on metallic oxide surfaces is known following the Langmuir kinetics, typically resulting in the formation of a monolayer[36]. The relationship between the surface response ($\Delta I_s$) and the BSA concentration ($C$) is analyzed using the classical Langmuir isotherm model using the general Langmuir equation as Eq. (15)[37]:

$$q = \frac{QC}{C + K_{eq}} \quad (15)$$

where $q$ is the amount of adsorbed adsorbate, $Q$ is the maximum saturated adsorption amount, and $K_{eq}$ is the adsorption equilibrium constant for the Langmuir adsorption process. Given that the SPR response is proportional to the adsorbed mass within the evanescent field, we substituted $q$ with the measured intensity change $\Delta I_s$, and $Q$ with the saturated intensity response.

## ■ RESULTS AND DISCUSSION

The real time recorded SPR reflection curve is shown in Fig. 4(a). Here, the light intensity recorded as $I_{water}$ represents the intensity in the flow cell filled with DI water before the experiment began. At the initial introduction of saline, the reflection intensity exhibits a distinct step change to $I_0$. Subsequent injection of BSA solution further increased the reflection intensity, with its steady-state value exhibiting a linear increase trend as the BSA concentration rose. Notably, after flushing with the saline solution, the reflection intensity does not decrease to the initial value $I_0$. This indicates that BSA molecules adsorb well on the $Al_2O_3$ surface.

The steady-state reflected intensity after BSA solution and saline against the BSA concentration are plotted in Fig. 4(b), respectively. The extracted bulk response and surface response against the BSA concentration are plotted in Fig. 4(c). The surface response $\Delta I_s$ vs. BSA concentration is fitted with Langmuir formula and an excellent fitting is resulted with $R^2$ value of 0.9975. Notably, the $K_{ep}$ of BSA is calculated as 13.38 $\mu M^{-1}$, which is close to the value reported (~10.3 $\mu M^{-1}$)[37]. This quantitative consistency reinforces that the adsorption phenomenon is attributable to interactions with BSA, ultimately form a dense monolayer at saturation. It confirms this procedure has the capability to accurately describe the saturation kinetics of BSA adsorption, making it a reasonable basis for calculating the RI of the adsorbed layer. The bulk responses in Fig. 4(c) exhibit a good linear fitting. This demonstrates that the RI of BSA solutions is in the linear of SPR sensing. The RI changes for varied BSA concentration solutions are calculated with the sensitivity determined by water and saline and shown in Fig. 5(a) as orange solid curve.

Based on the field-weighted model, different weights for 2 nm, 3 nm, 4 nm, and 5 nm thick surface layer, are used based on the simulated $L_{pd}$. Calculated RI of the bulk solution and the surface layer are shown in Fig. 5(a) and (b), respectively. The obtained bulk RI from the field-weighted model for each BSA concentration are compared to the results from Abbe refractometer, too. It is noticed that different thick surface layer does not influence the weight ($w_b = 1 - w_s$) significantly for the bulk solution as expected. This confirms the validity of neglecting the surface weighting term in the theoretical derivation for the bulk RI characterization. As the BSA concentration increased, discrepancy can be observed between SPR strategy and Abbe refractometer. We attribute the different here primarily to the fact that the Abbe refractometer's sensing substrate is $SiO_2$. The BSA molecule also exhibits strong adsorption properties on $SiO_2$ surfaces[32]. Only a drop of BSA solution is used in the Abbe refractometer measurement, and this adsorption reduces the BSA concentration effectively, leading to an underestimated RI value. Abbe refractometer uses a broadband incident light and this could also introduce considerable error when compared to SPR scheme that uses a single wavelength to sense RI changes.

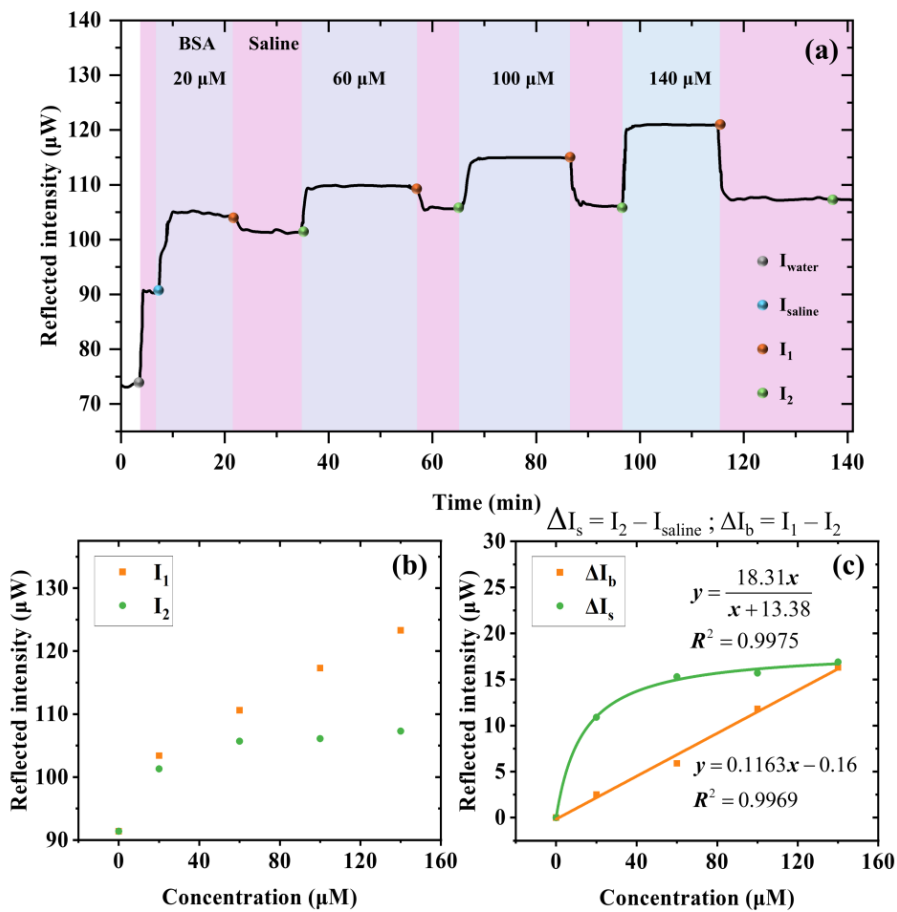

**Fig. 4.** (a) Time-resolved SPR reflected intensity response during sequential injection of DI water, saline solution, and BSA solution at different concentrations. (b) Reflected intensity extracted from the time traces versus BSA concentration. (c) Separated bulk and surface response change versus BSA concentration. Langmuir fit is used for the surface response and linear fit is used for the bulk response.

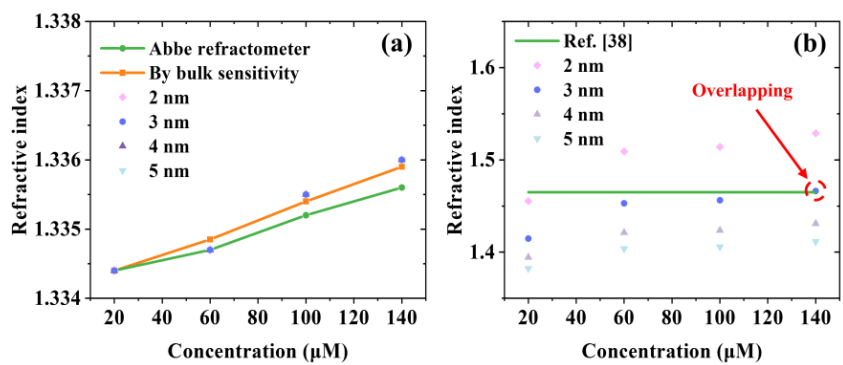

**Fig. 5.** (a) Comparison among bulk RI results from Abbe measurement, SPR intensity interrogation using bulk sensitivity, and field-weighted model calculation by using 2 nm – 5 nm thick single layer. (b) Comparison of BSA monolayer RI reported in literature [38] with calculated surface RI by using 2 nm – 5 nm thick single layer.

With field-weighted model, RI of the surface BSA layer with different thicknesses are shown in **Fig. 5(b)**. Unlike the bulk response, the surface RI exhibits a pronounced nonlinear dependence on BSA concentration. It increases rapidly in the low-concentration region and gradually approaches saturation in the high-concentration region. This trend aligns well with the Langmuir adsorption isotherm. Moreover, the $d_s$ have the significant impact on the surface layer RI calculation. As $d_s$ increases, the obtained surface RI gets lower. When $d_s$ = 3 nm, the RI of the surface layer reaches 1.4662 at the saturation adsorption, matching the 1.465 for a dense single layer of BSA reported in previous literature[38]. This result also verifies that the BSA adsorption layer in the experimental saturation condition approximates an ideal monolayer of approximately 3 nm.

Replacing the concentration of BSA with the RI values of each BSA solution as the *x*-axis values, the RI value of water is calculated in a backward manner. Compared to the saline solution, the reflected intensity change from water is -21 μW which corresponds to RI at 1.3321. This value has a difference of 0.0003 compared to the result of the Abbe measurement within the uncertainty range of the Abbe refractometer. This demonstrates that without using the sensitivity, the bulk RI can be extracted using the field-weighted model. This is also proved the observation in Fig. 5(a). The results from the sensitivity strategy and field-model match well. It should be emphasized that the sensitivity is not the goal of this research since SPR techniques have been demonstrated the sensing capabilities for tiny changes happening on the metal surface. Actually, a moderate sensitivity is enough and the stability of the SPR system is the most important for a sensitive, stable and reproducible measurements.

## ■ CONCLUSION

This study proposes and validates an SPR intensity interrogation sensing scheme combined with a field-weighted model based on the nature of the evanescent field distribution. With a single-channel intensity readout, the bulk and surface responses are extracted simultaneously and RI values of the bulk medium and surface layer are determined. The experimental extracted surface component aligns well with the Langmuir adsorption model. For the bulk RI derivation, we are confident that the results from the field-weighted model are more accurate compared to the measurements from an Abbe refractometer, with deviations at the $10^{-4}$ order of magnitude. Meanwhile, when BSA adsorption reached a steady state, the calculated RI of the surface layer is 1.4662 based on a weight using 3 nm thickness. These two parameters both matches reported values from literatures. These results demonstrate that this SPR sensing strategy can be used for surface adsorbed layer studies which are severe important in biological and medical research areas. Without any pre-knowledge of the surface layer, a single point data cannot be used to extract the thickness and the RI value of the surface layer. Although some information is known before sensing for most of the applications, a characterization strategy can provide thickness and RI value of the surface layer simultaneously without pre-knowledge is preferred. By extending the field-weighted model to multiple excitation wavelengths based on the fact that Al films support SPR in a broad range of wavelength, we envision establishing a system that decouples the RI and thickness of the surface layer, thereby further enhancing the analytical depth of this compact sensing platform.

## ■ AUTHOR INFORMATION

Corresponding Author

**Xue Han** - School of Optoelectrical Engineering and Instrumentation Science, Dalian University of Technology, Dalian 116024, China; orcid.org/0000-0001-9525-1176; Email: xue_han@dlut.edu.cn

Authors

**Zhiying Chen** – School of Optoelectrical Engineering and Instrumentation Science, Dalian University of Technology, Dalian 116024, China

**Zihao Luo** – School of Optoelectrical Engineering and Instrumentation Science, Dalian University of Technology, Dalian 116024, China

**Changsen Sun** – School of Optoelectrical Engineering and Instrumentation Science, Dalian University of Technology, Dalian 116024, China

**Dmitry V. Kiesewetter** – Peter the Great St. Petersburg Polytechnic University, St. Petersburg, Russia

**Sergey Krivosheev** – Peter the Great St. Petersburg Polytechnic University, St. Petersburg, Russia

**Sergey Magazinov** – Peter the Great St. Petersburg Polytechnic University, St. Petersburg, Russia

**Victor Malyugin** – Peter the Great St. Petersburg Polytechnic University, St. Petersburg, Russia

Author Contributions

The manuscript was written through contributions of all authors. All authors have given approval to the final version of the manuscript.

Notes

The authors declare no competing financial interest.

## ■ ACKNOWLEDGMENTS

The authors acknowledge the financial support from the National Key Research and Development Program of China (2024YFE0213500), the National Natural Science Foundation of China (Grant Nos. 62105053), Instrumental Analysis Center at Dalian University of Technology.

■ REFERENCES

(1) Knoll, W. Interfaces and thin films as seen by bound electromagnetic waves. Annu. Rev. Phys. Chem. 1998, 49, 569–638.

(2) Jodaylami, M. H.; Masson, J. F. Surface plasmon resonance sensing. Nat. Rev. Methods Primers 2025, 5 (1), 47.

(3) Dostálek, J.; Huang, C. J.; Knoll, W. In surface design: applications in bioscience and nanotechnology; Förch, R., et al., Eds.; Wiley-VCH: 2009; pp 29–53.

(4) Homola, J. Surface plasmon resonance sensors for detection of chemical and biological species. Chem. Rev. 2008, 108, 462–493.

(5) Mukhopadhyay, R. Surface plasmon resonance instruments diversify. Anal. Chem. 2005, 77, 313A–317A.

(6) Marek, P.; Homola, J. Surface plasmon resonance (SPR) sensors: approaching their limits? Opt. Express 2009, 17 (19), 16505–16517.

(7) Matsunaga, R.; Ujiie, K.; Inagaki, M.; Inagaki, I.; Fernández Pérez, J.; Yasuda, Y.; Mimasu, S.; Soga, S.; Tsumoto, K. High-throughput analysis system of interaction kinetics for data-driven antibody design. Sci. Rep. 2023, 13, 19417.

(8) Shi, X.; Kuai, L.; Xu, D.; Qiao, Y.; Chen, Y.; Di, B.; Xu, P. Surface plasmon resonance (SPR) for the binding kinetics analysis of synthetic cannabinoids: advancing CB1 receptor interaction studies. Int. J. Mol. Sci. 2025, 26 (8), 3692.

(9) Das, S.; Devireddy, R.; Gartia, M. R. Surface plasmon resonance (SPR) sensor for cancer biomarker detection. Biosensors 2023, 13, 396.

(10) Chen, T.; Xin, J.; Chang, S. J.; Chen, C. J.; Liu, J. T. Surface plasmon resonance (SPR) combined technology: a powerful tool for investigating interface phenomena. Adv. Mater. Interfaces 2023, 10 (8), 2202202.

(11) Chen, T.; Xin, J.; Chang, S. J.; Chen, C. J.; Liu, J. T. Surface plasmon resonance (SPR) combined technology: a powerful tool for investigating interface phenomena. Adv. Mater. Interfaces 2023, 10 (8), 2202202.

(12) Homola, J. In Surface plasmon resonance based sensors; Homola, J., Ed.; Springer: 2006; pp 3–44.

(13) Nakamoto, K.; Kurita, R.; Niwa, O. One-chip biosensor for simultaneous disease marker/calibration substance measurement in human urine by electrochemical surface plasmon resonance method. Biosens. Bioelectron. 2010, 26, 1536–1542.

(14) Fan, M.; Thompson, M.; Andrade, M. L.; Brolo, A. G. Silver nanoparticles on a plastic platform for localized surface plasmon resonance biosensing. Anal. Chem. 2010, 82, 6350–6352.

(15) Ferhan, A. R.; Jackman, J. A.; Sut, T. N.; Ho, N. J. Quantitative comparison of protein adsorption and conformational changes on dielectric-coated nanoplasmonic sensing arrays. Sensors 2018, 18 (4), 1283.

(16) Treebupachatsakul, T.; Shinnakerdchoke, S.; Pechprasarn, S. Sensing mechanisms of rough plasmonic surfaces for protein binding of surface plasmon resonance detection. Sensors 2023, 23 (7), 3377.

(17) Yesudasu, V.; Pradhan, H. S. Performance enhancement of a novel surface plasmon resonance biosensor using thallium bromide. IEEE Trans. Nano Biosci. 2022, 21 (2), 206–215.

(18) Olaya, C. M.; Hayazawa, N.; Hermosa, N.; Tanaka, T. Angular Goos–Hänchen shift sensor using a gold film enhanced by surface plasmon resonance. J. Phys. Chem. A 2021, 125 (1), 451–458.

(19) Michel, D.; Xiao, F.; Alameh, K. A compact, flexible fiber-optic surface plasmon resonance sensor with changeable sensor chips. Sens. Actuators B: Chem. 2017, 246, 258–261.

(20) Li, L.; Zhao, J.; Feng, N.; Liu, Y. Deep-learning-assisted wedge fiber optic surface plasmon resonance sensor for accuracy detection of trace mercury ions. Photon. Res. 2026, 14 (1), 40.

(21) Lee, J. S.; Huynh, T.; Lee, S. Y.; Lee, K. G.; Lee, J.; Tame, M.; Rockstuhl, C.; Lee, C. Quantum noise reduction in intensity-sensitive surface plasmon resonance sensors. Phys. Rev. A 2017, 96 (3), 033833.

(22) Chinowsky, T. M.; Strong, A.; Bartholomew, D.; Jorgensen Soelberg, S.; Notides, T.; Furlong, C.; Yee, S. S. Improving surface plasmon resonance sensor performance using critical-angle compensation. Proc. SPIE 1999, 104.

(23) Chinowsky, M. T.; Yee, S. S. Data analysis and calibration for a bulk-refractive-index-compensated surface plasmon resonance affinity sensor. Proc. SPIE 2001, 4578, 442.

(24) Svirelis, J.; Andersson, J.; Stradner, A.; Dahlin, A. Accurate correction of the "bulk response" in surface plasmon resonance sensing provides new insights on interactions involving lysozyme and poly (ethylene glycol). ACS Sens. 2022, 7 (4), 1175–1182.

(25) Zeng, B.; Gao, Y.; Bartoli, F. J. Differentiating surface and bulk interactions in nanoplasmonic interferometric sensor arrays. Nanoscale 2015, 7, 166.

(26) Liu, F.; Zhang, X.; Li, K.; Guo, T.; Ianoul, A.; Albert, J. Discrimination of bulk and surface refractive index change in plasmonic sensors with narrow bandwidth resonance combs. ACS Sens. 2021, 6 (8), 3013–3023.

(27) Ouberai, M. M.; Xu, K. R.; Welland, M. E. Effect of the interplay between protein and surface on the properties of adsorbed protein layers. Biomaterials 2014, 35 (24), 6157–6163.

(28) Vörös, J. The density and refractive index of adsorbing protein layers. Biophys. J. 2004, 87 (1), 553–561.

(29) Li, H.; Chang, J.; Hou, T.; Li, F. HRP-mimicking DNAzyme-catalyzed in situ generation of polyaniline to assist signal amplification for ultrasensitive surface plasmon resonance biosensing. Anal. Chem. 2017, 89, 673–680.

(30) Lambert, A. S.; Valiulis, S. N.; Malinick, A. S.; Tanabe, I.; Cheng, Q. Plasmonic biosensing with aluminum thin films under the Kretschmann configuration. Anal. Chem. 2020, 92 (13), 8654–8659.

(31) He, C. C.; Li, Y. H.; Yang, Y. X.; Fan, H. K.; Li, D. W.; Han, X. Sensitive aluminum SPR sensors prepared by thermal evaporation deposition. ACS Omega 2023, 8 (45), 43188–43196.

(32) Fukuzaki, S.; Urano, H.; Nagata, K. Adsorption of bovine serum albumin onto metal oxide surfaces. J. Ferment. Bioeng. 1996, 81 (2), 163–167.

(33) Otto, A. Excitation of nonradiative surface plasma waves in silver by the method of frustrated total reflection. Z. Physik 1968, 216, 398–410.

(34) Azzam, R. M. A.; Bashara, N. M. Ellipsometry and Polarized Light; North-Holland: 1987; p 464.

(35) Seitz, R.; Brings, R.; Geiger, R. Protein adsorption on solid–liquid interfaces monitored by laser-ellipsometry. Appl. Surf. Sci. 2005, 252 (1), 154–157.

(36) Bukackova, M.; Marsalek, R. Interaction of BSA with ZnO, TiO2, and CeO2 nanoparticles. Biophys. Chem. 2020, 267, 106475.

(37) Latour, R. A. The Langmuir isotherm: a commonly applied but misleading approach for the analysis of protein adsorption behavior. J. Biomed. Mater. Res. Part A 2015, 103 (3), 949–958.

(38) Kurrat, R. Adsorption of biomolecules on titanium oxide layers in biological model solutions. Dissertation, Zurich, 1998.